\documentclass[11pt,a4]{article}
\pdfoutput=1

\catcode`\@=11
\@addtoreset{equation}{section}

\RequirePackage{amsmath}
\RequirePackage{amssymb}
\RequirePackage{graphicx}

\RequirePackage[T1]{fontenc}
\RequirePackage[utf8]{inputenc}

\RequirePackage{mathrsfs}
\RequirePackage[sc]{mathpazo}
\RequirePackage{microtype}
\RequirePackage{wasysym}

\newcommand{\di}{\ensuremath{\mathrm{d}}}
\newcommand{\du}[2]{\ensuremath{}_{#1}^{\phantom{#1}{#2}}}
\newcommand{\ud}[2]{\ensuremath{}^{#1}_{\phantom{#1}{#2}}}
\DeclareMathOperator{\Tr}{Tr}
\newcommand{\SU}{\ensuremath{\mathrm{SU}}}
\newcommand{\U}{\ensuremath{\mathrm{U}}}

\normalsize
\divide\@tempdima\baselineskip
\@tempcnta=\@tempdima
\skip\@mpfootins = \skip\footins
\RequirePackage[numbers,sort&compress]{natbib}

\renewenvironment{thebibliography}[1]{%
\begin{oldthebibliography}{#1}%
\small%
\raggedright%
\setlength{\itemsep}{5pt}%
}%
{%
\end{oldthebibliography}%
}

\let\oldbibitem=\bibitem
\renewcommand{\bibitem}{%
  \filbreak
  \oldbibitem
}

\RequirePackage{tikz}

\allowdisplaybreaks

\begin{document}

\begin{flushright}
\parbox{4.2cm}
{KUNS--2328 \\ 
IPMU11--0054}
\end{flushright}

\vspace*{2cm}

\begin{center} {\Large 
    \textbf{Yangian symmetry in deformed WZNW models \\
    on squashed spheres}}
\vspace*{2cm}\\
{\large \textbf{Io Kawaguchi}$^{\dagger}$, 
  \textbf{Domenico Orlando}$^{\ast}$
  and 
  \textbf{Kentaroh Yoshida}$^{\dagger}$
}
\end{center}
\vspace*{1cm}
\begin{center}
  $^{\dagger}${\it Department of Physics, Kyoto University \\ 
    Kyoto 606-8502, Japan} 
  \vspace*{0.5cm} \\ 
  $^{\ast}${\it Institute for the Physics and Mathematics of the Universe (IPMU), \\
    The University of Tokyo, 
    Kashiwa, Chiba 277-8568, Japan. 
  }
\end{center}

\vspace{4cm}

\begin{abstract}
  We introduce a deformation of the Wess--Zumino--Novikov--Witten
  model with three-dimensional squashed sphere target space. We show
  how with an appropriate choice of Wess--Zumino and boundary terms it
  is possible to construct an infinite family of conserved charges
  realizing an $\SU(2)$ Yangian. Finally we discuss the running of the
  squashing parameter under renormalization group flow.

\end{abstract}

\thispagestyle{empty}
\setcounter{page}{0}

\newpage 

\section{Introduction}

In 1978, L\"uscher and Pohlmeyer found that the $\mathrm{O}(N)$
non-linear sigma model in two dimensions has a hidden
infinite-dimensional symmetry~\cite{Luscher} (for related work,
see~\cite{Bernard,MacKay} and the comprehensive textbook~\cite{AAR}).
This symmetry received a mathematical formulation in the work of
Drinfel'd~\cite{Drinfeld}, who called it Yangian.  A similar
construction can be extended to non-linear sigma models on symmetric
spaces where it is again possible to construct an infinite number of
non-local charges which realize a Yangian.  An important point to note
here is that a flat conserved current exists always from which
an infinite number of non-local charges can be generated, for example
by following the treatment in~\cite{BIZZ}. Since AdS spaces and round spheres are
symmetric spaces, the same argument can be used in the AdS/CFT
correspondence~\cite{Maldacena} as pointed out in~\cite{Mandal,BPR}
and extensively discussed in~\cite{Zarembo}, where a list of symmetric
spaces with potential AdS/CFT application is given.

\medskip 

While the symmetric cases have been studied intensively and are well
understood, finding hidden infinite-dimensional symmetries for
non-symmetric cosets is still an open problem. These are of physical
interest and have been intensively studied in recent years. Typical
examples are the Schr\"odinger and Lifshitz
spacetimes~\cite{Son,BM,KLM,SYY} that appear in applications of the
AdS/CFT correspondence to condensed matter physics.  Other examples of
non-symmetric cosets are three-dimensional squashed spheres and warped
AdS spaces.  In topologically massive gravity~\cite{TMG} for example,
warped AdS spaces appear as classical solutions~\cite{warped}.  They
also appear as the near-horizon geometry of the extremal Kerr black
hole~\cite{BH,Kerr/CFT}, and play an important role in the Kerr/CFT
correspondence~\cite{Kerr/CFT}.  In addition, squashed spheres and
warped AdS spaces appear in string theory contexts such as
deformations of heterotic string
backgrounds~\cite{Hetero,Hetero-two,S2,S3,Orlando:2006cc,S4},
T--dualities of some string backgrounds~\cite{OU}, as well as in
relation to condensed matter system~\cite{Kraus}.

\medskip 

Recently, it has been shown that the Yangian symmetry is preserved 
in a non-linear sigma model on a three-dimensional 
squashed sphere~\cite{KY}:  
\begin{equation}
  \di s^2 = \frac{L^2}{4}\left[\, \di \theta^2 + \cos^2{\theta}\,\di \phi^2 
    +\left( 1 + C \right) \left( \di \psi +\sin{\theta}\,\di \phi\right)^2\, \right]\,. \label{sphere}
\end{equation}
This geometry is a deformation of a three-dimensional round sphere
$S^3$, where the constant parameter $C$ measures how much the original
round sphere is squashed. The round sphere is realized for $C=0$ and
its isometry is $\mathrm{SO}(4)=\SU(2)_{\rm L}\times \SU(2)_{\rm
  R}$\,.  For $C\neq 0$, the isometry is reduced to $\SU(2)_{\rm
  L}\times \U(1)_{\rm R}$\,.  Recall that the presence of a flat
conserved current is crucial for the realization of the Yangian
symmetry.  In the case discussed in~\cite{KY} it has been shown that
the flat conserved current can be constructed by improving the Noether
current only for $C\geq 0$.  On the other hand, the positivity of $C$
might seem strange since the preferred range of $C$ is $-1 \leq C \leq
0$ in physical setups such as the deformation of the heterotic string
background discussed in~\cite{Hetero}.  Also in the recent T--duality
argument on how to realize a Yangian symmetry with a squashed sphere
target space~\cite{ORU}, the range of $C$ is naturally restricted to
$-1 \leq C \leq 0$\,.  It is therefore an interesting question to
consider how the Yangian symmetry can be realized in this range.

Here we argue that a Wess--Zumino (\textsc{wz}) term should be added
to the sigma model action. 
The theory with the \textsc{wz} term can be called the squashed
Wess--Zumino--Novikov--Witten (\textsc{wznw}) model. Our main result
is that for an appropriate choice of the normalization of the
\textsc{wz} term (which is constrained by dimensionality ad quantum
consistency) and adding a boundary term it is in general possible to
construct a flat conserved current. This is equivalent to saying that the
Yangian algebra can be
realized 
even for $-1\leq C \leq 0$\,.

\medskip 

This letter is organized as follows. In Section 2 we introduce the
action of the \textsc{wznw} model on a three-dimensional squashed
sphere. Then we show that a flat conserved current can be constructed
i) for a certain value of the coefficient of the \textsc{wz} term and ii) for
general values by improving the Noether current. In Section 3 the
Yangian algebra is computed with the standard Poisson bracket. In
Section 4 the renormalization group flow is discussed by computing
one-loop $\beta$-functions.  Section 5 is devoted to conclusion and
discussion.

\section{The flatness condition for the squashed \textsc{wznw} model}

Consider the action of a two-dimensional sigma model with squashed
sphere target space:
\newcommand{\SSM}{\ensuremath{S_{\sigma\text{\textsc{m}}}}}
\begin{equation}
  \SSM = -\frac{1}{2\lambda^2}\int\!\!\!\int\!\! \di t\di x\!\left[
    (\partial_{\mu}\theta)^2 + \cos^2\theta\,(\partial_{\mu}\phi)^2  
    + (1+C)(\partial_{\mu}\psi + \sin\theta\,\partial_{\mu}\phi)^2
  \right]\,. 
\end{equation}
The base space is a two-dimensional Minkowski spacetime 
with coordinates $x^{\mu}=(t,x)$ and metric $\eta_{\mu\nu}=(-1,+1)$\,. 
The parameter $\lambda^2$ is the bare coupling constant. 

\medskip 

It is convenient to introduce the $\SU(2)_{\rm L}$ group element
\begin{equation}
  g = {\rm e}^{\phi T_1}{\rm e}^{\theta T_2} {\rm e}^{\psi T_3}\,, \qquad g \in \SU(2)_{\rm L}\,,
\end{equation}
where the $\SU(2)_{\rm L}$ generators $T_A$~$(A=1,2,3)$ satisfy the relations 
\begin{equation}
  [T_A,T_B] = \epsilon\du{AB}{C} T_C\,, \qquad \Tr(T_{A}T_{B})=-\frac{1}{2}\delta_{AB}\,,
\end{equation}
where $\epsilon \du{AB}{C} $ is the anti-symmetric tensor. 
By using the left-invariant current $J_{\mu}$ on $\SU(2)_{\rm L}$ given by 
\begin{equation}
  J_{\mu}=g^{-1}\partial_{\mu}g\,, 
\end{equation}
the sigma model action is rewritten as 
\begin{equation}
  \SSM = \frac{1}{\lambda^{2}}\int\!\!\!\int\!\!\di t\di x\,
  \eta^{\mu\nu} \left[ \Tr(J_{\mu}J_{\nu}) - 2 C \Tr
  \left(T_{3}J_{\mu}\right)\Tr\left(T_{3}J_{\nu}\right) \right]
  \,. 
  \label{action}
\end{equation}
\medskip 
Next let us introduce the Wess-Zumino (\textsc{wz}) term, 
\newcommand{\SWZ}{\ensuremath{S_{\text{\textsc{wz}}}}}
\begin{equation}
  \SWZ \equiv  \frac{n}{12\pi}
  \int_{0}^{1}\!\!\! \di s\! \int\!\!\!\int\!\! \di t\di x\, 
  \epsilon_{\hat{\mu}\hat{\nu}\hat{\rho}}\Tr(J^{\hat{\mu}}_{s}J^{\hat{\nu}}_{s}J^{\hat{\rho}}_{s})
  \qquad (n\in \mathbb{Z})\,,
\end{equation}
where $n$ is an integer.  It is given by a three-form defined on
a fictitious three-dimensional base space with coordinates
$x^{\hat{\mu}}=(x^{\mu},s)$\,. The anti-symmetric tensor
$\epsilon_{\hat{\mu}\hat{\nu}\hat{\rho}}$ is normalized as
$\epsilon_{txs}=+1$\,.  The coordinate $s$ describes an interval, $s
\in [0,1]$\,. The variable $g(x^{\mu},s) \equiv g_s(x^{\mu})$
interpolates between the unit element and $g(x)$, \emph{i.e.} $g_{0}(x)=1$
and $g_{1}(x)=g(x)$\,.  The current $J_s^{\hat{\mu}}$ is defined as
$J_s^{\hat{\mu}} \equiv g_s^{-1}\partial^{\hat{\mu}}g_s$\,.

\medskip 

Note that the \textsc{wz} term is the same as in the usual \textsc{wznw} model 
because of dimensionality and quantum consistency.
In terms of the angle variables, the integrand of the \textsc{wz} term 
is proportional to  the volume form, 
\begin{equation}
  \epsilon_{\hat{\mu}\hat{\nu}\hat{\rho}}\Tr(J^{\hat{\mu}}_{s}J^{\hat{\nu}}_{s}J^{\hat{\rho}}_{s})\,  
  \di t\,\di x\,\di s  
  ~~\propto~~
  \epsilon_{\hat{\mu}\hat{\nu}\hat{\rho}}\cos\theta_s\,
  \partial^{\hat{\mu}}\phi_s\,\partial^{\hat{\nu}}\theta_s\,\partial^{\hat{\rho}}\psi_s\,
  \di t\,\di x\,\di s\,.
\end{equation}
This is simply due to dimensionality. The target space is now three-dimensional 
and so the three-form to define the \textsc{wz} term must be proportional to the volume form. 
Thus the form of the \textsc{wz} term is essentially fixed by the dimensionality in the present case, 
up to an overall constant. This constant has to be discretized in the standard way. 
At the end of the day, the \textsc{wz} term turns out to be exactly the same as in the $\SU(2)$ \textsc{wznw} model. 

\medskip 

The model described by the sum of $\SSM$ and $\SWZ$ can be called the
squashed Wess--Zumino--Novikov--Witten (\textsc{wznw}) model. The
resulting action is given by
\newcommand{\SSQ}{\ensuremath{S_{\text{Sq}}}}
\begin{align}
  \SSQ &=  \SSM + \SWZ \, ,  \label{eq:S-SqWZNW}\\
  \SSM &= \frac{1}{\lambda^{2}}\int\!\!\!\int\!\!\di t\di x\,
  \eta^{\mu\nu}\Bigl[\Tr(J_{\mu}J_{\nu}) - 2 C \Tr
  \left(T_{3}J_{\mu}\right)\Tr\left(T_{3}J_{\nu}\right) \Bigr]\,,
  \label{lambda} \\
  \SWZ &=
  \frac{n}{12\pi}\int_{0}^{1}\!\!\!ds\!\!\int\!\!\!\int\!\!\di t\di
  x\, \epsilon_{\hat{\mu}\hat{\nu}\hat{\rho}} \Tr \left(J_{s}^{\hat{\mu}}J_{s}^{\hat{\nu}}J_{s}^{\hat{\rho}}\right),
  \qquad (n\in{\mathbb Z})\,.
\end{align}
%
The corresponding equations of motion have the form
\begin{equation}
  \partial_{\mu}J^{\mu}-2C\,\Tr(T_{3}\partial_{\mu}J^{\mu})T_{3} 
  + 2C\,\Tr(T_{3}J^{\mu})[T_{3},J_{\mu}]-\frac{K}{2}\epsilon^{\mu\nu}[J_{\mu},J_{\nu}]=0\,,
\end{equation}
where $J^\mu = \eta^{\mu \nu} J_\nu$ and 
\begin{equation}
  K \equiv \frac{n\lambda^2}{8\pi}\,.
\end{equation}
Note that $C=0$, $K=1$ is the fixed point for the standard
\textsc{wznw} model.

\medskip 

For $C\neq 0$, the Lagrangian has $\SU(2)_{\rm L}\times \U(1)_{\rm R}$
symmetry.  For the $\SU(2)_{\rm L}$ part, the Noether current $j_{\mu}$
is given by
\begin{equation}
j_{\mu}= \partial_{\mu}g\cdot g^{-1} -2C\,\Tr(T_{3}J_{\mu})\,gT_{3}g^{-1} 
- K\,\epsilon_{\mu\nu}\,\partial^{\nu}g\cdot g^{-1}\,.
\end{equation}
In terms of the current, the equations of motion take the form
\begin{equation}
  g^{-1} \partial_\mu j^\mu g = 0  \, ,
\end{equation}
or, equivalently
\begin{equation}
  \di  * j = 0 \, .  
\end{equation}
This shows that as usual, the Noether current is defined up to the
Hodge dual of an exact $1$-form. Hence we can add an improvement term
$I_{\mu}$:
\begin{equation}
  j_\mu \mapsto j_\mu + I_\mu \, ; \hspace{2em} \, I_{\mu} \equiv A\, \epsilon_{\mu\nu}\partial^{\nu}f\,, 
  \label{improve}
\end{equation}
where $A$ is an arbitrary constant and $f$ a scalar function. As
pointed out in~\cite{KY}, the function $f$ must have the form
\begin{equation}
  f=g\,T_{3}\,g^{-1} \label{fn}
\end{equation}
in order to satisfy the flatness condition. This improvement term
corresponds to a boundary term of the form
\begin{equation}
\label{eq:S-boundary}
  S_{\text{\textsc{bdry}}} = \frac{A}{\lambda^2} \int \di t \di x \, \epsilon_{\mu \nu} \partial^\mu J_3^\nu\,.
\end{equation}


The resulting improved current $\tilde{j}_{\mu}$ depends on the three
parameters $C$, $K$ and $A$ and is given by
\begin{equation}
  \tilde{j}_{\mu} = j_{\mu} + I_{\mu} = \partial_{\mu}g\cdot g^{-1} - 2 C \Tr(T_{3}J_{\mu})\,gT_{3}g^{-1} -
  K\,\epsilon_{\mu\nu}\,\partial^{\nu}g\cdot g^{-1} + A\,
  \epsilon_{\mu\nu}\partial^{\nu}(gT_{3}g^{-1})\,.
\end{equation}

\medskip 

In order to define an infinite family of conserved charges we require
the improved current $\tilde{j}$ to be flat. Because of the $\SU(2)$
symmetry, the flatness condition reduces to a single condition on the
three parameters:
\begin{equation}
  \epsilon^{\mu\nu}(\partial_{\mu}\tilde{j}_{\nu} - \tilde{j}_{\mu}\tilde{j}_{\nu})
  =\left(C-\frac{CK^2}{1+C}-A^2\right)\epsilon_{\mu\nu}\Tr(T_{3}[J^{\mu},J^{\nu}]) g T_{3} g^{-1} = 0\,. 
\label{flat}
\end{equation}

It is interesting to consider separately the two cases $A=0$ and $A
\neq 0$, with and without the improvement term:
\begin{enumerate}
\item For $A=0$, supposing that $C \geq -1$\,, we obtain the condition
  \begin{equation}
    C = K^2 - 1 \, .
  \end{equation}
  Thus, as opposed to the case without a \textsc{wz} term~\cite{KY}, a flat
  conserved current can be constructed even for negative values of
  $C$. This condition is reminiscent of the one discussed
  in~\cite{Hetero}, but it is not exactly the same since the model at
  hand is not conformal.
\item For $A\neq 0$, it is possible to construct a flat conserved
  current if $A$ is taken to be
  \begin{equation}
    A^2= C\left(1-\frac{K^2}{1+C}\right)\,. 
    \label{A-cond}
  \end{equation}
  In this expression we have to restrict the range of $C$ to $C >
  -1$\,.  Since $K$ is non-vanishing, a negative value of $C$ is
  possible.  The equation is of second order in $C$ and can be solved
  algebraically. The solutions are given by
  \begin{equation}
    \label{eq:C-function-A}
    C^{\pm}_A (\lambda) = \frac{1}{2}\left[A^2-1 + K^2\pm 
    \sqrt{4 A^2 + \left(A^2-1 + K^2\right)^2} \right] \, .
  \end{equation}
\end{enumerate}

In the following we will keep the three constant parameters $C$\,, $K$
and $A$\,, but assume either of the above conditions to be satisfied.
When referring to the current, we will always mean the improved
current and omit the tilde for simplicity.

\section{The Yangian algebra}

So far we have constructed a flat conserved current for $\SU(2)_{\rm
  L}$ by imposing appropriate conditions on the parameters.  Given
this, it is always possible to construct an infinite number of
non-local conserved charges by following the prescription
in~\cite{BIZZ}.
For example, the Noether charge $Q^{A}_{(0)}$ and the first non-local
charge $Q^{A}_{(1)}$ are given by
\begin{align}
  Q_{(0)}^{A} &\equiv \int\!\! \di x\, j_{t}^{A}(x)\,, \\
  Q_{(1)}^{A} &\equiv  - \int\!\! \di x\, j_{x}^{A}(x) + \frac{1}{4}\int\!\!\!\int\!\! \di x\di y\, 
  \epsilon(x-y)\,\epsilon \du{BC}{A}\,
  j_{t}^{B}(x)j_{t}^{C}(y)\,, 
\end{align}
where
\begin{equation}
  \epsilon(x - y) =
  \begin{cases}
    +1 & \text{if $ x > y $} \\
    -1 & \text{if $ x < y $}.
  \end{cases}
\end{equation}
The conservation law for $Q_{(1)}^A$ is a consequence of the flatness
condition.

\medskip 

The Poisson brackets of the charges can be found via the current
algebra.  Imposing the flatness condition in Eq.~\eqref{A-cond}, the
currents form an algebra in terms of the standard Poisson bracket for the
dynamical variables. It reads:
\begin{subequations}
  \label{current}
  \begin{align}
    \{ j_{t}^{A}(x),j_{t}^{B}(y) \}_{\rm P} &= \epsilon
    \ud{AB}{C}\, j_{t}^{C}(x)\, \delta(x-y)\,
    - 2K\delta^{AB}\partial_{x}\delta(x-y)\,, \\
    \{ j_{t}^{A}(x),j_{x}^{B}(y) \}_{\rm P} &= \epsilon\ud{AB}{C}\,
    j_{x}^{C}(x)\, \delta(x-y) + \left(1+C+\frac{K^2}{1+C}\right)
    \delta^{AB}\partial_{x}\delta(x-y)\,, \\
    \begin{split}
      \{ j_{x}^{A}(x),j_{x}^{B}(y) \}_{\rm P} &= -
      \left(C+\frac{K^2}{1+C}\right)
      \epsilon \ud{AB}{C}\, j_{t}^{C}(x)\, \delta(x-y)\, - 2K
      \epsilon \ud{AB}{C}\,
      j_{x}^{C}(x)\, \delta(x-y) \,  \\
      & \phantom{={ }} - 2K\delta^{AB}\partial_{x}\delta(x-y)\,.
    \end{split}
  \end{align}
\end{subequations}
Note that the result of~\cite{KY} is reproduced for $K=0$ (no
\textsc{wz} term).  For $C=0$ (no squashing) and $K=1$, the
right-moving current $j_{\rm R} \equiv (j_t + j_x)/2$ vanishes
(\emph{i.e.}  $j_t = - j_x$) and only the algebra for the left-moving
current $j_{\rm L} \equiv (j_t - j_x)/2=j_t$ remains.

\medskip 

The current algebra (\ref{current}) leads to the following (Yangian)
algebra for the charges:
\begin{subequations}
  \begin{align}
    \{ Q_{(0)}^{A},Q_{(0)}^{B} \}_{\rm P} &= \epsilon\ud{AB}{C}\,Q_{(0)}^{C}\,, \\
    \{ Q_{(0)}^{A},Q_{(1)}^{B} \}_{\rm P} &= \epsilon\ud{AB}{C}\,Q_{(1)}^{C}\,, \\
    \{ Q_{(1)}^{A},Q_{(1)}^{B} \}_{\rm P} &= \epsilon\ud{AB}{C}
    \left[Q_{(2)}^{C} + \frac{1}{12}Q_{(0)}^{C}(Q_{(0)})^2 + 2K
      Q_{(1)}^{C}
      -\left(C+\frac{K^2}{1+C}\right) Q_{(0)}^{C}\right]\,,
  \end{align}
\end{subequations}
where $Q_{(2)}^{A}$ is the second non-local charge defined as
\begin{multline}
  Q_{(2)}^{A} \equiv \frac{1}{12}\int\!\!\!\int\!\!\!\int\!\!
  \di x\,\di y\,\di z\, \epsilon(x-y)\,
  \epsilon(y-z)\,\delta_{BC}\,
  \left[\,j_{t}^{A}(x)\,j_{t}^{B}(y)\,j_{t}^{C}(z)
    - j_{t}^{B}(x)\,j_{t}^{A}(y)\,j_{t}^{C}(z)\,\right] \\
   + \frac{1}{2}\int\!\!\!\int\!\! \di x\di y\,
  \epsilon(x-y)\,\epsilon\du{BC}{A}\,
  j_{t}^{B}(x)\,\,j_{x}^{C}(y) \,. 
\end{multline}
It is straightforward to check that the $\SU(2)_{\rm L}$ Serre
relations are satisfied:
\begin{subequations}
  \begin{align}
    \{ \{ Q_{(1)}^{+},Q_{(1)}^{-} \}_{\rm P} , Q_{(1)}^{3} \}_{\rm P}
    &=
    \frac{1}{4}Q_{(0)}^{3}(Q_{(1)}^{+}Q_{(0)}^{-}-Q_{(1)}^{-}Q_{(0)}^{+})\,, \\
    \{ \{ Q_{(1)}^{3},Q_{(1)}^{\pm} \}_{\rm P} , Q_{(1)}^{\pm} \}_{\rm
      P} &= \frac{1}{4} Q_{(0)}^{\pm} (Q_{(1)}^{3} Q_{(0)}^{\pm} -
    Q_{(1)}^{\pm} Q_{(0)}^{3})\,, \\
    \begin{split}
      \{ \{ Q_{(1)}^{+},Q_{(1)}^{-} \}_{\rm P} , Q_{(1)}^{\pm} \}_{\rm
        P} &\pm 2 \{ \{ Q_{(1)}^{3},Q_{(1)}^{\pm} \}_{\rm P} ,
      Q_{(1)}^{3} \}_{\rm P} \\
      &
      =\frac{1}{4}Q_{(0)}^{\pm}(Q_{(1)}^{+}Q_{(0)}^{-}-Q_{(1)}^{-}Q_{(0)}^{+})
      \pm
      \frac{1}{2}Q_{(0)}^{3}(Q_{(1)}^{3}Q_{(0)}^{\pm}-Q_{(1)}^{\pm}Q_{(0)}^{3})\,,
    \end{split}
  \end{align}
\end{subequations}
where 
\begin{equation}
  j_{\mu}^{\pm} \equiv j_{\mu}^1 \pm i j_{\mu}^2\,. 
\end{equation}
Thus in the squashed \textsc{wznw} model, the $\SU(2)_{\rm L}$ Yangian
symmetry is realized as an infinite-dimensional symmetry.

In fact it is possible to rewrite the Yangian algebra in a standard
form by using the fact that higher charges are defined up to $Q_{(0)}$
and shifting the first and second charges as follows:
\begin{align}
  {\widetilde Q}_{(1)} &= Q_{(1)} - K \, Q_{(0)} \, ,  \\
  {\widetilde Q}_{(2)} &= Q_{(2)} - A^2 \, Q_{(0)} \, .
\end{align}

\section{The $\beta$-function of the squashed \textsc{wznw} model}

The model described by the action $\SSQ$ in Eq.~\eqref{eq:S-SqWZNW} is
in general not conformal even though, as we have shown in the previous
section, it preserves a Yangian symmetry. In this section we study the
running of the squashing parameter $C$ and the coupling constant
$\lambda$ and discuss the relation between the condition obtained in
the previous section and the renormalization group (RG) flow.


As it is already the case for the standard \textsc{wznw} model, the
qualitative behavior can be understood semiclassically at one-loop
level\footnote{The \textsc{rg}~flow is the
  Bianchi~\textsc{ix}~Ricci~flow in the axisymmetric case with a
  \textsc{wz}~term. As such it was already analyzed
  in~\cite{Bakas:2006bz,Orlando:2007cc}.}. Consider the action in
Eq.~(\ref{lambda}).  First decompose the $\SU(2)$ group element $g$
into a classical solution $g_{0}$ and a quantum fluctuation $\xi$ as
follows:
\begin{equation}
  g = g_{0}\,\mathrm{e}^{\lambda\xi} \hspace{2em} \xi \in \mathrm{Lie}[\SU(2)] \, ;
\end{equation}
then expand the left-invariant current $J$ with respect to
$\lambda$\,,
\begin{equation}
  \begin{split}
    J &= g^{-1} \di g \\
    &= {\rm e}^{-\lambda\xi}(g_{0}^{-1} \di g_{0}){\rm e}^{\lambda\xi}
    + \lambda\int_{0}^{1}\!\!\!\di t\,{\rm e}^{-t\lambda\xi}(\di \xi){\rm e}^{t\lambda\xi} \\
    &= J_{0} + \lambda\left(\di \xi +\left[J_{0},\xi \right] \right) +
    \frac{\lambda^{2}}{2} \left(-\left[\xi, \di \xi
      \right]+\left[\left[J_{0},\xi\right],\xi\right]\right)
    + {\mathcal O}\left(\lambda^{3}\right)\,,
  \end{split}
\end{equation}
where $J_{0} = g_{0}^{-1} \di g_{0}$\,.
By rewriting the fluctuation $\xi$ as 
\begin{equation}
  \xi = \xi^{1}T_1 + \xi^{2}T_2 + \frac{1}{\sqrt{1+C}}\xi^{3}T_3
\end{equation}
and introducing the new vector notation
$\vec{\xi}=(\xi^1,\xi^2,\xi^3)$\,, the action at second order in the
fluctuation is given by
\begin{equation}
  S_{\rm quad}=\int\!\!\!\int\!\! \di t \di x\,
\left[-\frac{1}{2}\eta^{\mu\nu}\partial_{\mu}\vec{\xi}\cdot\partial_{\nu}\vec{\xi} 
+ \vec{J_{0}}_{\mu}\cdot\vec{v}^{\mu} 
+ \frac{C}{2}\vec{J_{0}}_{\mu}{\mathsf M}^{\mu\nu}\vec{J_{0}}_{\nu} \right]\,,
\end{equation}
where the following quantities have been introduced: 
\begin{align}
  \vec{J_{0}}_{\mu} &=
  \begin{pmatrix}
    J_{0\mu}^{1} \\ J_{0\mu}^{2} \\J_{0\mu}^{3}
  \end{pmatrix}
  \,, \\ 
  \vec{v}_{\mu} &=
  \begin{pmatrix}
    -\frac{1}{2}\left(\sqrt{1+C}\eta_{\mu\nu}-\frac{\lambda^{2}n}{8\pi\sqrt{1+C}}\epsilon_{\mu\nu}\right)\left(\xi^{2}\partial^{\nu}\xi^{3}-\xi^{3}\partial^{\nu}\xi^{2}\right)  \\
    \frac{1}{2}\left(\sqrt{1+C}\eta_{\mu\nu}-\frac{\lambda^{2}n}{8\pi\sqrt{1+C}}\epsilon_{\mu\nu}\right)\left(\xi^{1}\partial^{\nu}\xi^{3}-\xi^{3}\partial^{\nu}\xi^{1}\right)\\
    -\frac{1}{2}\left((1-C)\eta_{\mu\nu}-\frac{\lambda^{2}n}{8\pi}\epsilon_{\mu\nu}\right)\left(\xi^{1}\partial^{\nu}\xi^{2}-\xi^{2}\partial^{\nu}\xi^{1}\right)
  \end{pmatrix}
  \,,\\
  %
%
%
  \begin{split} {\mathsf M}_{\mu\nu} &= \eta_{\mu \nu}
    \begin{pmatrix}
      -\xi^{2}\xi^{2} & \xi^{1}\xi^{2} &  -\frac{1}{2}\sqrt{1+C} \, \xi^{1}\xi^{3}\\
      \xi^{1}\xi^{2} & -\xi^{1}\xi^{1} & -\frac{1}{2}\sqrt{1+C} \, \xi^{2}\xi^{3}\\
      -\frac{1}{2} \sqrt{1+C} \, \xi^{1}\xi^{3} & -\frac{1}{2}
      \sqrt{1+C} \, \xi^{2}\xi^{3} & \xi^{1}\xi^{1} + \xi^{2}\xi^{2}
    \end{pmatrix} + \\
    & + \frac{\lambda^{2} \, n \, \epsilon_{\mu\nu}}{16\pi\sqrt{1+C}}
    \begin{pmatrix}
      0 & 0 & \xi^1 \xi^3 \\
      0 & 0 & \xi^2 \xi^3 \\
      \xi^1 \xi^3 & \xi^2 \xi^3 & 0
    \end{pmatrix} \, .
  \end{split}
\end{align}
The divergent part in the one-loop effective action can be evaluated as follows:
\begin{eqnarray}
  {\rm e}^{iW\left[g_{0}\right]}
  &=& 
  {\rm e}^{ i \SSQ [g_{0}]}
  \!\int\left[\di \xi\right]
  {\rm e}^{iS_{\rm quad}\left[g_{0};\xi\right]}
  \nonumber \\
  &=& 
  {\rm e}^{i \SSQ [g_{0}]}
  \!\int[\di \xi]
  \exp\left\{-\frac{i}{2}\int\!\!\di^{2}x \, \eta^{\mu\nu} \partial_{\mu}\vec{\xi}\cdot\partial_{\nu}\vec{\xi}
  \right\} \nonumber \\
  &&\times\left[
    1+i\!\int\!\!\di^{2}x\vec{J_{0}}_{\mu}\!\!\cdot\!\vec{v}^{\mu}(x)-\frac{1}{2}\int\!\!\di^{2}x\!\int\!\!\di^{2}y\vec{J_{0}}_{\mu}\!\!\cdot\!\vec{v}^{\mu}(x)\vec{J_{0}}_{\nu}\!\!\cdot\!\vec{v}^{\nu}(y)\right. \nonumber \\
  &&\left.+i\frac{C}{2}\int\!\!\di^{2}x\vec{J_{0}}_{\mu}{\mathsf M}^{\mu\nu}\vec{J}_{0\nu}(x) + {\mathcal O}\left(\xi^{3}\right)\right] \nonumber \\
  &=& {\rm e}^{i \SSQ[g_{0}]} \nonumber \\
  &&\times\left\{1+\frac{i}{16\pi}\left[1-C-\frac{1}{1+C}\left(\frac{\lambda^{2}n}{8\pi}\right)^{2}\right]\log{\frac{\Lambda^{2}}{\mu^{2}}}\int\!\!\di^{2}x\eta^{\mu\nu}\left[J_{0\mu}^{1}J_{0\nu}^{1}+J_{0\mu}^{2}J_{0\nu}^{2}\right] \right. \nonumber \\
&&\left. +\frac{i}{16\pi}\left[\left(1+C\right)^{2}-\left(\frac{\lambda^{2}n}{8\pi}\right)^{2}\right]\log{\frac{\Lambda^{2}}{\mu^{2}}}\int\!\!\di^{2}x\eta^{\mu\nu}J_{0\mu}^{3}J_{0\nu}^{3}
  + \text{finite}~\right\}\,, 
\end{eqnarray}
where we have used the two-point function:
\begin{equation}
  \left<\xi^{A}(x)\xi^{B}(y)\right> = -\int\!\frac{\di^{2}k}{(2\pi)^{2}}\,
  {\rm e}^{-ik\cdot(x-y)}\frac{1}{k^{2}}\,. 
\end{equation}
From the explicit expression we can read off the one-loop effective
coupling constant and the squashing parameter:
\begin{align}
  \lambda_{\rm R}^2 &= \lambda^2 + \frac{\lambda^{4}}{8\pi}\left(1-C-\frac{1}{1+C}\left(\frac{\lambda^{2}n}{8\pi}\right)^{2}\right)\log\frac{\Lambda^{2}}{\mu^{2}}\,, \\
  C_{\rm R} &= C -
  \frac{\lambda^{2}}{4\pi}C(1+C)\log\frac{\Lambda^{2}}{\mu^{2}}\,,
\end{align}
where $\Lambda$ and $\mu$ are the ultraviolet (UV) and the infrared
(IR) cut-off, respectively. By deriving with respect to $\mu$ we
obtain the one-loop $\beta$-functions for $\lambda_{\rm R}^2$ and
$C_{\rm R}$:
\begin{align}
  \mu\frac{\partial\lambda_{\rm R}^{2}}{\partial\mu} &= -\frac{\lambda_{\rm R}^{4}}{4\pi}
  \left\{1-C_{\rm R}-\frac{1}{1+C_{\rm R}}\left(\frac{\lambda_{\rm R}^{2}n}{8\pi}\right)^{2}
  \right\}\,, \label{RG1} \\
  \mu\frac{\partial C_{\rm R}}{\partial\mu} &= \frac{\lambda_{\rm R}^{2}}{2\pi}
  C_{\rm R}\left(1+C_{\rm R}\right)\,. \label{RG2}
\end{align}
Note that the coefficient of the \textsc{wz} term is quantized since $\SU(2)$ is
a compact group.

\medskip 

Before discussing the RG flow described by (\ref{RG1}) and
(\ref{RG2})\,, let us recall the condition we obtained from the
flatness of the $\SU(2)_{\rm L}$ Noether current. By writing $K$
explicitly in terms of the coupling constant $\lambda$\,, the flatness
condition for the conserved current in Eq.~\eqref{A-cond} can be rewritten
as
\begin{equation}
  C-\frac{C}{1+C}\left(\frac{n\lambda^2}{8\pi}\right)^2 - A^2 =0\,,
\label{curve}
\end{equation}
and $C$ can be expressed as a function of $\lambda$ for fixed values
of $A$ as in Eq.~\eqref{eq:C-function-A}:
\begin{equation}
  C^{\pm}_A (\lambda) = \frac{1}{2}\left[(A^2-1) + \left(\frac{n\lambda^2}{8\pi}\right)^2\right] \pm 
  \sqrt{A^2 + \frac{1}{4}\left(A^2-1 + \left(\frac{n\lambda^2}{8\pi}\right)^2\right)^2}
  \,.
\end{equation}

\medskip 

It is interesting to discuss the RG flow with respect to the flatness
condition (\ref{curve}). A typical example ($n=20$) is represented in
Fig.~\ref{RG:fig}. There is a unique IR fixed point on the line $C=0$
for $\lambda_R = 8 \pi /n $, which is the same as in the usual
$\SU(2)$ \textsc{wznw} model. As $n$ increases, the fixed point
approaches $\lambda^2=0$ on the line $C=0$\,. It is worth noting that
the critical surface is defined by $C > -1$ and the universality class
is characterized by the unique IR fixed point.  In other words, the
flow remains in the same universality class as the $\SU(2)$
\textsc{wzw} model.  The locus $C=-1$ deserves some special
attention. As already shown in~\cite{Hetero}, it corresponds to a
decompactification limit in which the squashed three-sphere
degenerates to the direct product $S^2 \times S^1$. In this case, the
action $\SSM$ in Eq.~\eqref{action} describes the well-studied
non-linear sigma model on the coset $\SU(2) / \U(1) \sim S^2$, also known as the
$\mathrm{O}(3) $ model.

\medskip 

The $(\lambda^2, C)$ plane is separated into four regions by red lines
corresponding to $C = C^{\pm}_0 (\lambda)$ (this is the locus where
the flatness condition is satisfied without improvement term
$I_\mu$). In these regions, $A^2 \gtrless 0$. The green line in
Fig.~\ref{RG:fig} corresponds to the locus $\{ C^\pm_1 \}$
(\emph{i.e.}~$A^2 = 1$), while the cyan lines represent $\{
C^\pm_{-i/2} \}$ (\emph{i.e.}~$A^2=-1/4$)\,.  Those are typical
examples and, by varying the value of $A^2$\,, the green and cyan
lines cover all of either regions.

\medskip 

Up to this point we have implicitly considered $A^2 > 0$ since we
interpreted it as resulting from the boundary term contribution in
Eq.~\eqref{eq:S-boundary}. The RG flow analysis suggests that this
condition is too strong and that it is possible to analytically continue to
$A^2 < 0$. In fact, this is natural from the point of view of the
$\mathrm{O}(3)$ sigma model at $C = - 1$ (which is by itself the
analytical continuation of our Lorentzian worldsheet) where the
boundary contribution is interpreted as a Hopf (instanton)
term. Moreover, the conserved charges depend only on $A^2$ and no
imaginary term is generated by the continuation.


\tikzstyle{whitebox}=[draw=black, fill=white, text=black]

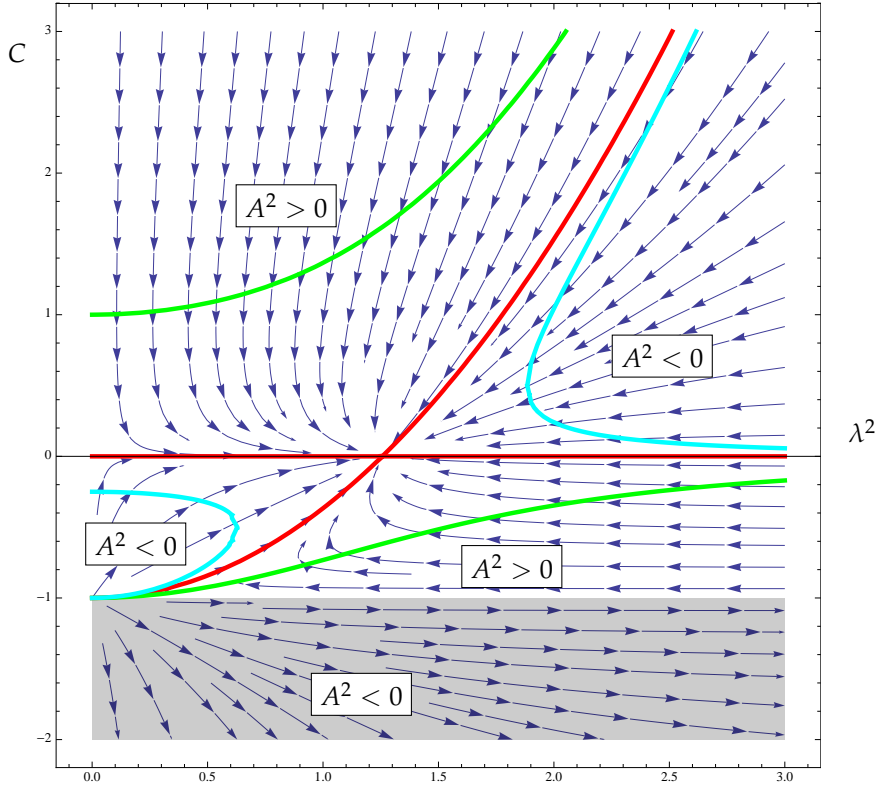
\begin{figure}
  \centering
  \begin{tikzpicture}
    \node [inner sep=0pt,above right] {\includegraphics[scale=.7]{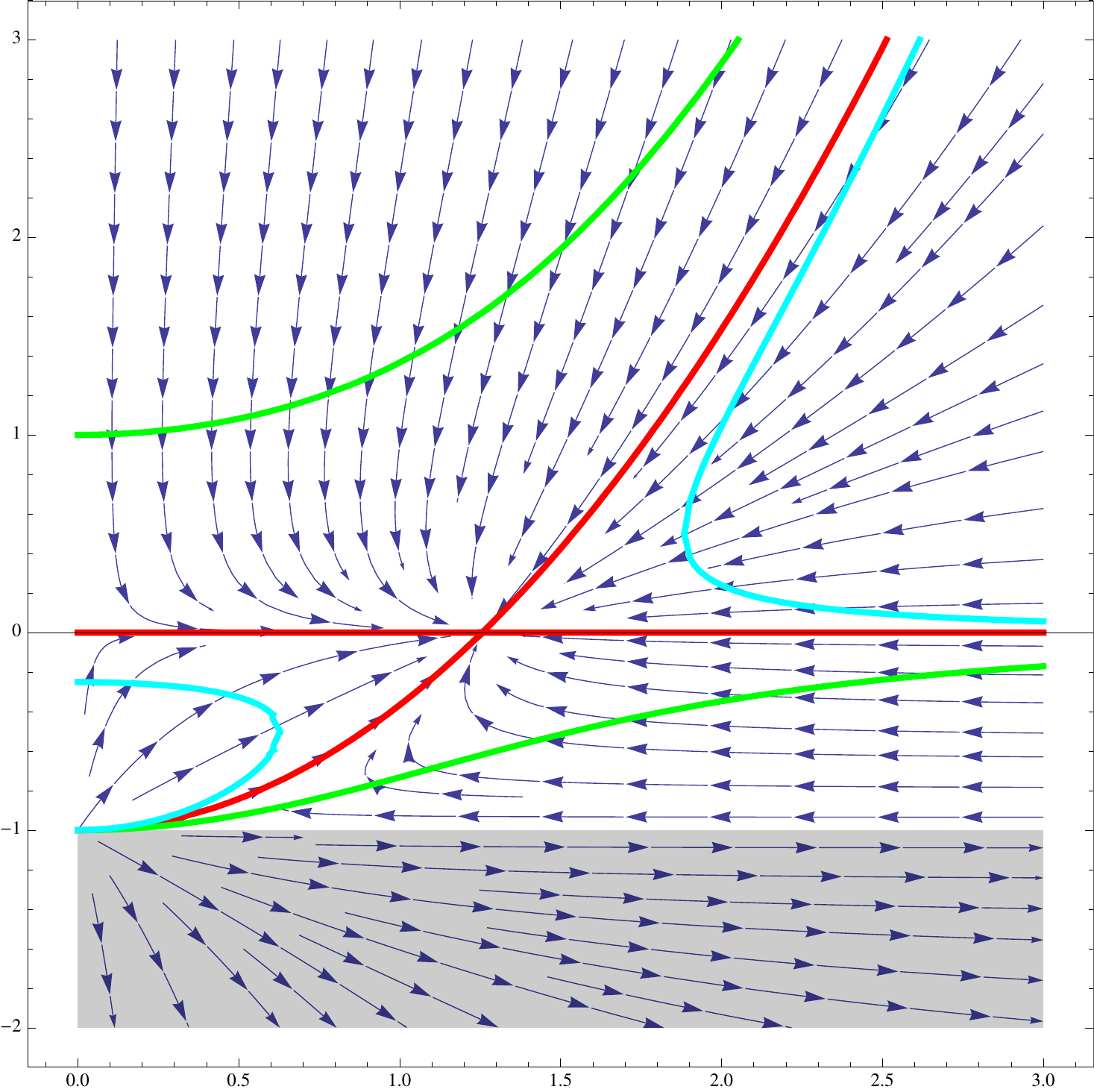}};
    \begin{small}
      \draw (4,8) node[anchor=north east,whitebox]{$A^2>0$};
      \draw (7,3.2) node[anchor=north east,whitebox]{$A^2>0$};
      \draw (9,6) node[anchor=north east,whitebox]{$A^2<0$};
      \draw (2,3.5) node[anchor=north east,whitebox]{$A^2<0$};
      \draw (5,1.5) node[anchor=north east,whitebox]{$A^2<0$};
      \draw (0,10) node[anchor=north east]{$C$};
      \draw (11.3,5) node[anchor=north east]{$\lambda^2$};
    \end{small}
  \end{tikzpicture}
  \begin{small}
    \caption{RG flow for $n=20$\,. The vertical axis shows $C$ and the
      horizontal one $\lambda^2$\,. The unique IR fixed point on
      the line $C=0$ is the undeformed $\SU(2)$ \textsc{wznw}
      model. The red line depicts $C_\pm[0]$, the green line $C_\pm[1]$,
      the cyan line $C_\pm [-i/2]$. The flow is not defined in the
      grey region $C<-1$.}
  \end{small}
  \label{RG:fig}
\end{figure}

\newpage

\section{Conclusion and Discussion}

In this letter we have discussed the Yangian symmetry of the squashed
\textsc{wznw} model.  The \textsc{wz} term is fixed by dimensionality up to an overall
coefficient
. For a special value of this coefficient, a flat
conserved current can be constructed directly.  For general values, it
is possible to construct a flat conserved current using a current
improvement term. 

\medskip 


Although we have discussed only the case of the squashed sphere, it is
straightforward to carry out the same analysis for warped
$\mathrm{AdS}$ spaces via double Wick rotations. It would also be
interesting to consider a generalization of the present analysis to
higher-dimensional cases such as squashed $S^7$, possibly using
arguments similar to the ones given in~\cite{Hatsuda}.

\medskip

Another avenue of research is to include world-sheet fermions. For
symmetric spaces, the relation between world-sheet fermions and
non-local charges are well studied (for example, see~\cite{AF,AAR} and
 related work in~\cite{CZ,SZ,local2,susy}). Similarly, we expect in our
case that it is possible to find a flat conserved current
(and a Yangian algebra) by including appropriate world-sheet
fermions. Of course, it would be even more interesting to study
space-time fermions by introducing deformations of super Lie groups as
target spaces.

\medskip 

The next natural step consists in understanding the physical
implications of the Yangian symmetry that we have described. In
particular it will be interesting to see whether the existence of the
infinite charges implies complete integrability in the sense of
Liouville. In fact, since the squashed sigma model is thought to be a
continuum limit of the \textsc{xxz} model~\cite{FR} (even though the
Yangian symmetry is closely related to the \textsc{xxx} model), it is
possible that we have found a form of ``partial'' integrability,
related to a ``subclass'' of the soliton solutions in the squashed
sigma model.

\subsection*{Acknowledgments}
We would like to thank Y.~Hatsuda, H.~Kawai, T.~Okada and M.~Staudacher for
illuminating discussions and S.~Reffert for comments on the
manuscript.  This work was supported by the scientific grants from the
ministry of education, science, sports, and culture of Japan
(No.\,22740160), and in part by the Grant-in-Aid for the Global COE
Program ``The Next Generation of Physics, Spun from Universality and
Emergence'' from the Ministry of Education, Culture, Sports, Science
and Technology (MEXT) of Japan.  DO acknowledges support by the World
Premier International Research Center Initiative (WPI Initiative).





\end{document}